\documentclass{article}
\usepackage{arxiv}
\usepackage{graphicx}
\usepackage{tikz}
\usepackage{subcaption}
\usepackage{hyperref}
\hypersetup{colorlinks,allcolors=black}
\usepackage{natbib}
\usepackage{gensymb}
\usepackage{float}

\title{AIFS - ECMWF's data-driven forecasting system}
\author{Simon Lang\thanks{equal contribution} \And Mihai Alexe\footnotemark[1] \And Matthew Chantry \And Jesper Dramsch \And Florian Pinault \And Baudouin Raoult \And Mariana C. A. Clare \And Christian Lessig \And Michael Maier-Gerber \And Linus Magnusson \And Zied Ben Bouallègue \And Ana Prieto Nemesio \And Peter D. Dueben \And Andrew Brown \And Florian Pappenberger \And Florence Rabier}
\date{May 2024}

\begin{document}

\maketitle

\begin{abstract}
Machine learning-based weather forecasting models have quickly emerged as a promising methodology for accurate medium-range global weather forecasting. Here, we introduce the Artificial Intelligence Forecasting System (AIFS), a data driven forecast model developed by the European Centre for Medium-Range Weather Forecasts (ECMWF). AIFS is based on a graph neural network (GNN) encoder and decoder, and a sliding window transformer processor, and is trained on ECMWF's ERA5 re-analysis and ECMWF's operational numerical weather prediction (NWP) analyses. It has a flexible and modular design and supports several levels of parallelism to enable training on high-resolution input data. AIFS forecast skill is assessed by comparing its forecasts to NWP analyses and direct observational data. We show that AIFS produces highly skilled forecasts for upper-air variables, surface weather parameters and tropical cyclone tracks. AIFS is run four times daily alongside ECMWF's physics-based NWP model and forecasts are available to the public under ECMWF's open data policy.  
\end{abstract}

\section{Introduction}
Accurate weather forecasts are of great importance for society, helping to save lives and protect property. The last years have seen substantial progress in the realm of data-driven weather forecasting. Following exploratory work (\cite{Dueben2018}), researchers and technological companies like Google, Huawei and NVIDIA have built data-driven weather forecasting models that outperform leading physics-based global numerical weather prediction (NWP) models in many of the standard forecast scores (e.g. \cite{pathak2022fourcastnet, keisler2022forecasting, lam2022graphcast, Chen2023, bi2023accurate, nguyen2023scaling}, see also \cite{benbouallegue2023rise} for verification results), like geopotential at 500 hPa root mean squared error (RMSE) and anomaly correlation coefficient (ACC). At the same time, forecasts can be produced within minutes on a single GPU (graphics processing unit), compared to the O(1000) CPUs required for physics-based NWP forecasts.

The data-driven models are trained on historical weather data, usually a subset of the reanalysis dataset ERA5 (\cite{hersbach2020era5}) produced by the European Centre for Medium-Range Weather Forecasts (ECMWF) in the context of the EU Copernicus programme. When producing real-time forecasts, they currently rely on operational NWP initial conditions, such as ECMWF's 4D-Var (\cite{rabier4d}) analysis.

The decision was made at ECMWF to implement an experimental data-driven forecasting system ready for operational use. This involved the implementation of the data-driven forecast model, together with an end-to-end pipeline covering training and inference, including dataset generation, reproducible training, operational inference, the use of operational verification tools, product generation and dissemination of forecast data.

ECMWF’s data-driven forecast model is called AIFS – Artificial Intelligence Forecasting System, a homage to ECMWF’s IFS, the Integrated Forecasting System. AIFS is implemented in Python and utilises the PyTorch (\cite{paszke2019pytorch}) and PyTorch Geometric (\cite{fey2019fast}) machine learning frameworks. Together with AIFS we develop the Anemoi (Greek: 'Winds') framework, a toolbox that aims to provide building blocks for data driven weather forecast models.

\section{Model}
A diverse set of deep learning architectures is being employed in the context of data-driven weather forecasts, e.g., vision transformers (\cite{bi2023accurate,nguyen2023scaling}), neural operators (\cite{pathak2022fourcastnet}) and graph neural networks (GNNs, \cite{keisler2022forecasting,lam2022graphcast}).

\begin{figure}[htpb]
\centering
\begin{subfigure}{0.48\textwidth}
    \includegraphics[trim={0 0 420 0}, clip, width=\linewidth]{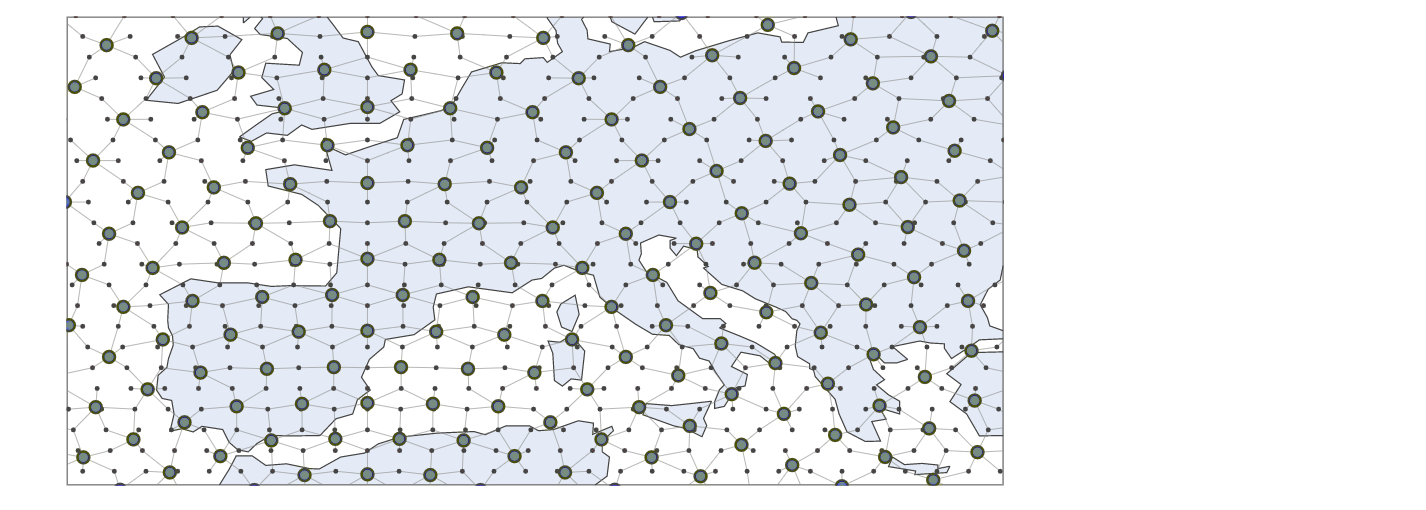}
\end{subfigure}
\hfill
\begin{subfigure}{0.48\textwidth}
    \includegraphics[trim={0 0 420 0}, clip, width=\linewidth]{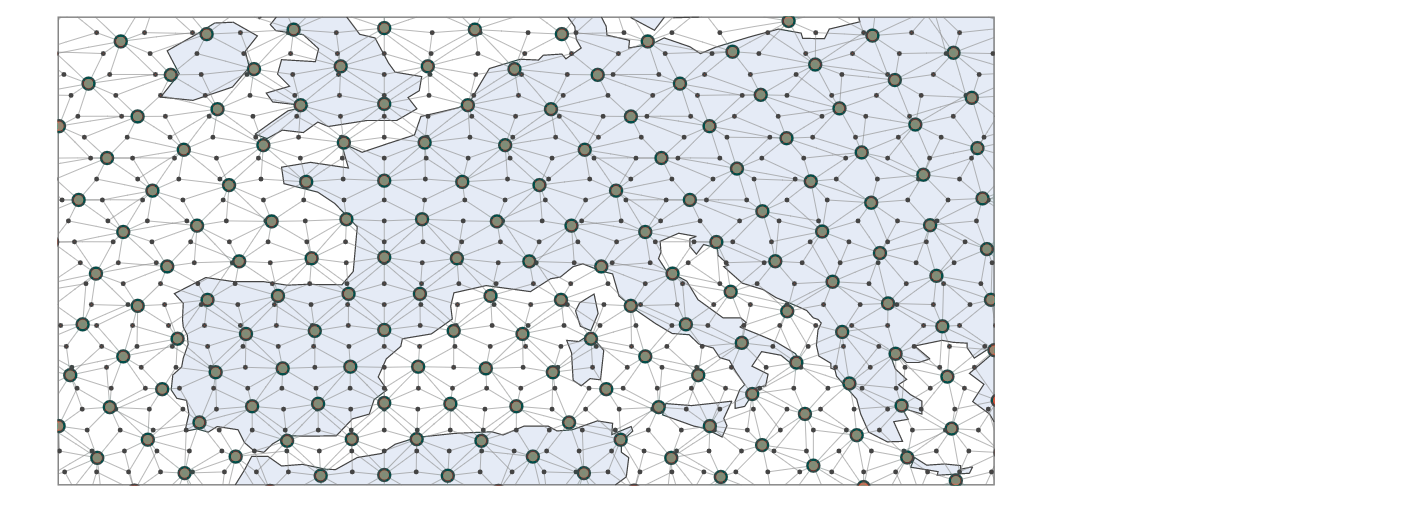}
\end{subfigure}
\caption{Example visualisation of an AIFS encoder (left) and decoder (right) graph. ERA5 grid points are shown as small black circles, processor grid points as larger grey circles. Edges are shown as black lines. Only a small region of the globe is plotted for visibility.}
\label{fig:encoder-decoder}
\end{figure}

The first version of AIFS was introduced in experimental operational mode in October 2023 (\cite{lang2024aifs}). Its architecture was inspired by \cite{keisler2022forecasting} and \cite{lam2022graphcast}. Specifically, it consisted of an encoder-processor-decoder design similar to that described in \cite{battaglia2016interaction}. The encoder aggregates information from the input data grid onto the lower resolution processor grid (Fig.~\ref{fig:encoder-decoder}). Here, a cut-off radius is employed with a certain distance around each processor grid point. The decoder projects the latent state back onto the input data grid. In the decoder, each grid point of the input data grid is connected to its three closest processor grid points. The edge features of the encoder and decoder graph are edge length and direction.

\begin{figure}[htpb]
\centering
\includegraphics[trim={220 100 220 100}, clip, width=\linewidth]{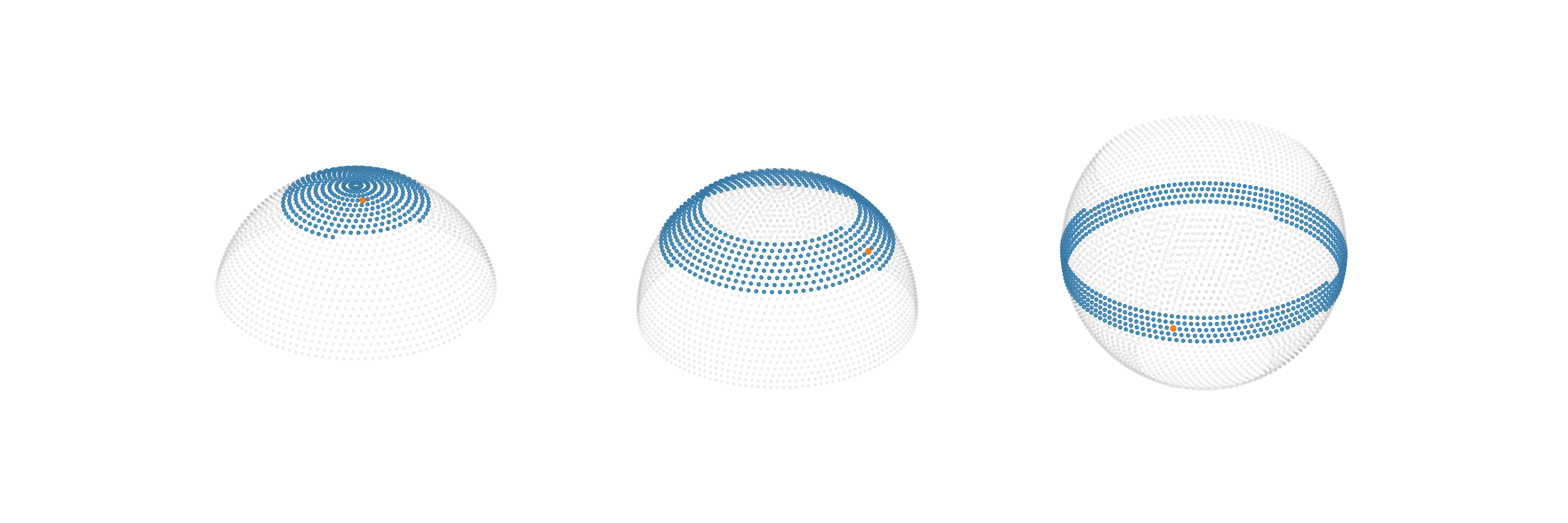}
\caption{Attention windows (grid points highlighted in blue) for different grid points (red). The grey grid points show an example of how far information can travel within 6 layers of the processor (AIFS has 16 processor layers in total). For illustration purposes, a lower-resolution grid is shown than that used in the AIFS.}
\label{fig:atten}
\end{figure}

\begin{figure}[htpb]
\centering
\begin{subfigure}{0.42\textwidth}
    \includegraphics[trim={50 415 300 70}, clip, width=\linewidth]{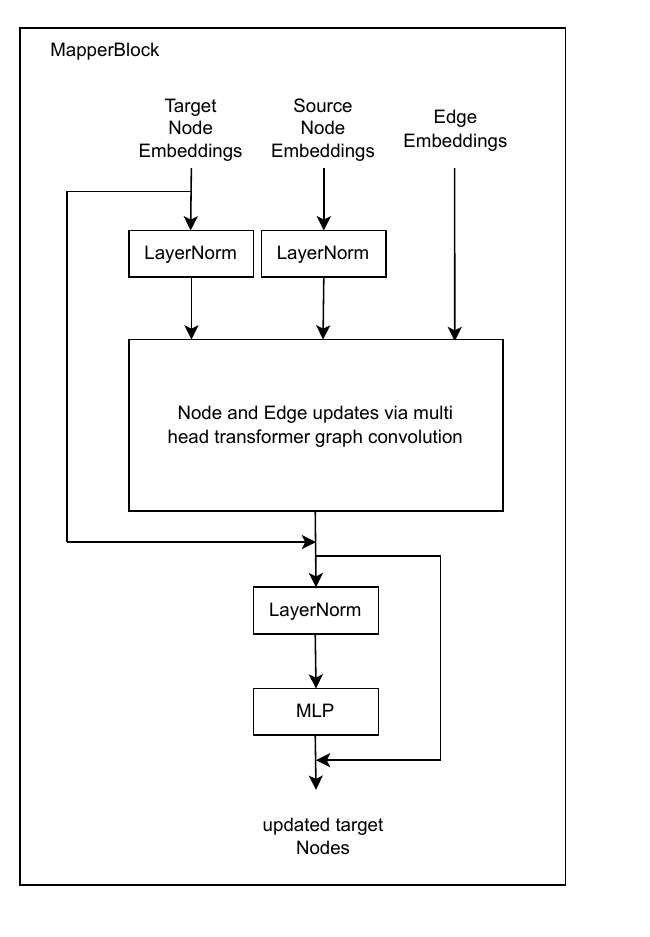}
\end{subfigure}
\begin{subfigure}{0.42\textwidth}
    \includegraphics[width=\linewidth]{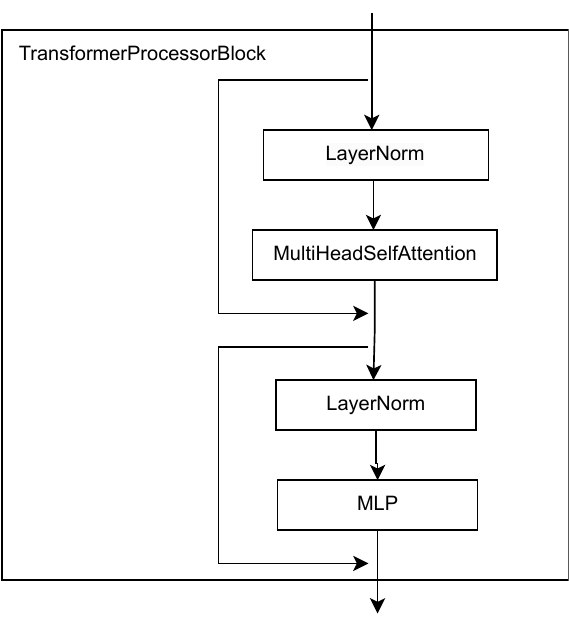}
\end{subfigure}
\caption{AIFS encoder / decoder and processor block schematics: GNN block (left), processor block (right). The GNN block uses a multi-head graph transformer convolution operation to update the nodes and the edges of the processor, whereas the \textit{pre-norm} transformer block relies on multi-head self-attention (\cite{vaswani2023attention}).}
\label{fig:layers}
\end{figure}

As in \cite{lam2022graphcast}, the hidden processor grid was based on an icosphere, though at a lower refinement level with approximately 10,000 nodes and 80,000 edges. The spatial resolution of the first AIFS version was approximately 1\degree. AIFS underwent an update in February 2024 to a spatial resolution of approximately 0.25\degree. The GNN flavour used in the encoder and decoder changed to an attention / transformer based variant following \cite{shi2021masked}. The processor is now based on a pre-norm transformer with shifted window attention (\cite{child2019generating}, \cite{beltagy2020longformer}, \cite{dao2023flashattention2}, \cite{jiang2023mistral}) and GELU activation function (\cite{hendrycks2023gaussian}). Consequently, it does not take explicit edge information into account but attention is instead computed along latitude bands (see Fig.~\ref{fig:atten}). Attention windows are chosen in such a way that a complete grid neighbourhood is always included. The processor grid is now based on a O96 octahedral reduced Gaussian grid (\cite{Wedi2014}), with 40320 grid points (approximately 1 degree spatial resolution, which is roughly the same as the processor resolution of \cite{lam2022graphcast}). The AIFS has 16 processor layers in total. Schematics of AIFS encoder-decoder and processor blocks are shown in Figure~\ref{fig:layers}.

The encoder and decoder operate on data at the native ERA5 spatial grid resolution, the N320 reduced Gaussian grid (approximately 31~km). This is in contrast to other data-driven weather models. One advantage of reduced Gaussian grids is that they have a much more uniform resolution around the globe, without grid spacing collapsing close to the poles. As a consequence, the total number of grid points and edges in the encoder and decoder is reduced compared to a regular longitude-latitude grid of similar resolution. The N320 reduced Gaussian grid has 542,080 grid points, while a 0.25\degree longitude-latitude grid has 1,038,240 grid points.

Input data and edge features are embedded via a linear layer before the encoder maps the input data from the ERA5 grid to the processor grid. The decoder maps the latent space back to the ERA5 input grid, before a linear layer projects it back to physical space.

Another feature of AIFS is that we add eight \textit{learnable} parameters (features) to each node on the input and the processor grids, as well as to the edges of the encoder and decoder graphs. This allows the network to learn additional useful features that are not directly present in the input data or pre-defined as either node or edge features.

To reduce the GPU memory footprint, AIFS makes extensive use of activation checkpointing during the forward model pass: instead of saving intermediate activations in GPU memory, these are re-computed during the backward pass (\cite{chen2016training}).

In addition to data parallelism, where a batch is divided into smaller sub-batches and these are processed simultaneously across multiple GPUs,
AIFS implements sequence / tensor parallelism. Here, one instance of the neural network is split across multiple GPUs (e.g. \cite{kurth2022fourcastnet}). This enables the training on high resolution data and over multiple auto-regressive steps. We support two modes of sequence parallelism. In the GNN attention encoder, decoder and in the transformer processor, attention heads are sharded across GPUs, with synchronisation relying on efficient all-to-all communication primitives, as proposed by \cite{jacobs2023deepspeed}. As an alternative option for the GNN encoder and decoder, we also implement node and edge sharding according to 1-hop graph neighbourhoods. Here, device-to-device communication is minimised through a judicious re-ordering of the graph nodes, such that only a single gather and reduce operation is required in the forward and backward pass of the model, respectively.

The benefits of tensor parallelism are twofold: the computational load can be distributed between GPUs to reduce run-time, and activations are split between GPUs, increasing the amount of GPU memory that is available to one model instance. The increased amount of available GPU memory makes training on high-resolution data possible. Weak parallel scaling tests have demonstrated that AIFS can scale quasi-linearly up to at least 2048 GPUs.

ECMWF's goal is to develop a probabilistic forecasting system. To this end, AIFS also implements a device-to-device communication pattern that allows for ensemble-based training towards probabilistic scores (e.g. \cite{doi:10.1198/016214506000001437}), such as fair scores (\cite{Ferro2013}). Complementary to the sequence parallelism described above, ensemble members can be distributed across several GPUs. This allows a direct probabilistic score-based optimisation of larger ensembles during training.

\section{Training}
AIFS is trained to produce 6-hour forecasts. It receives as input a representation of the atmospheric states (ERA5 or ECMWF's operational analysis) at $t_{-6h}$, $t_0$, and then forecasts the state at time $t_{+6h}$. The full list of input and output fields of AIFS is shown in Table~\ref{table:variables}. 
\begin{table}[htbp]
\centering
\begin{tabular}{|p{5cm}|p{3cm}|p{3cm}|}
\hline
\textbf{Field} & \textbf{Level type} & \textbf{Input/Output} \\ 
\hline
Geopotential, horizontal and vertical wind components, specific humidity, temperature & Pressure level: 50, 100, 150, 200, 250, 300, 400, 500, 600, 700, 850, 925, 1000 & Both \\ 
\hline
Surface pressure, mean sea-level pressure, skin temperature, 2 m temperature, 2 m dewpoint temperature, 10 m horizontal wind components, total column water & Surface & Both \\ 
\hline
Total precipitation, convective precipitation  & Surface & Output \\ 
\hline
Land-sea mask, orography, standard deviation of sub-grid orography, slope of sub-scale orography, insolation, latitude/longitude, time of day/day of year & Surface & Input \\ 
\hline
\end{tabular}
\caption{Input and output variables of AIFS.}
\label{table:variables}
\end{table}

Forecasts for longer lead-times are calculated in an auto-regressive fashion by initialising the model from its own prediction - so called \textit{rollout}. In order to improve forecast scores, rollout is also used during training (\cite{keisler2022forecasting}). Here, AIFS follows the training principles established by \cite{lam2022graphcast}: a \textit{pre-training} phase, during which the model is given the task to forecast 6~h ahead is followed by a second phase, where the model produces a forecast up to 72~h. Gradients flow through the entire forecast chain during backpropagation.

Pre-training was performed on ERA5 for the years 1979 to 2020 with a cosine learning rate (LR) schedule and a total of 260,000 steps. The LR is increased from 0 to $10^{-4}$ during the first 1000 steps, then it is annealed to a minimum of $3 \times 10^{-7}$. The pre-training is then followed by rollout on ERA5 for the years 1979 to 2018, this time with a LR of $6 \times 10^{-7}$. As in \cite{lam2022graphcast} we increase the rollout every 1000 training steps up to a maximum of 72~h (12 auto-regressive steps). Finally, to further improve forecast performance, we fine-tune the model on operational real-time IFS NWP analyses. This is done via another round of rollout training, this time using IFS operational analysis data from 2019 and 2020. For fine-tuning and initialisation of the model during inference, IFS fields are interpolated from their native O1280 resolution (approximately 0.1\degree) down to N320 (approximately 0.25\degree). As an optimizer, we use AdamW (\cite{loshchilov2018decoupled}) with the $\beta$-coefficients set to 0.9 and 0.95.

Input and output states are normalised to unit variance and zero mean for each level. Some of the forcing variables, like orography, are min-max normalised. The loss (objective) function is an area-weighted mean squared error (MSE) between the target atmospheric state and prediction. A loss scaling is applied for each output variable. The scaling was chosen empirically such that all prognostic variables have roughly equal contributions to the loss, with the exception of the vertical velocities, for which the weight was reduced. The loss weights also decrease linearly with height, which means that levels in the upper atmosphere (e.g., 50~hPa) contribute relatively little to the total loss value.

Data parallelism is used for training, with a batch size of 16. One model instance is split across four 40GB A100 GPUs within one node. Training is done using mixed precision (\cite{micikevicius2018mixed}), and the entire process takes about one week, with 64 GPUs in total. Running a 10~day forecast takes approximately 2 minutes 30 seconds on a single A100 GPU, including the input and output of forecast data.

\section{Results}
\begin{figure}[htbp]
\centering
    \includegraphics[width=\linewidth]{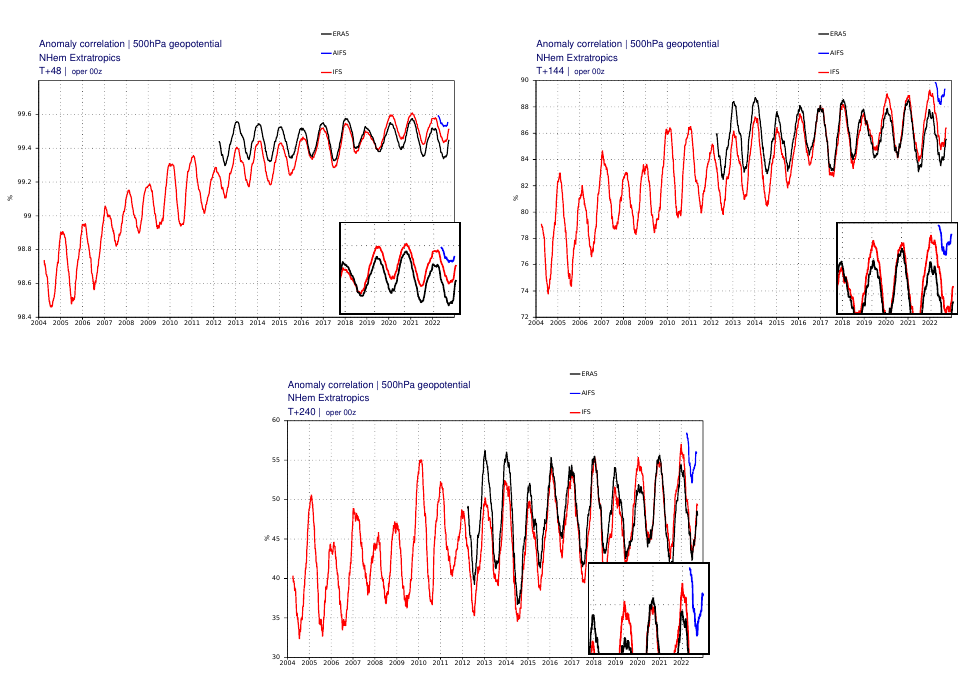}
\caption{Northern Hemisphere ACC (anomaly correlation) of geopotential at 500~hPa for AIFS (blue) and IFS (red) and ERA5 (black) forecasts for different years: 2 day forecasts (top left), 6 day forecasts (top right) and 10 day forecasts (bottom). Forecasts are initialised on 00 UTC each day and shown is a 30 day running mean. Insets show a zoomed-in view.}
\label{fig:ccaf}
\end{figure}

In general, AIFS produces highly skilful forecasts as can be seen from the ACC for the Northern Hemisphere compared to IFS and ERA5 forecasts (Fig.~\ref{fig:ccaf}), amounting to a forecast advantage of more than 12~h for longer lead-times. The time-series shows how the forecast skill of IFS evolved over the years. Here, improvements in the forecast model, data-assimilation system, observation usage, and the inherent variability of atmospheric predictability drive changes in skill. In comparison, ERA5 is a frozen system, and forecast skill changes are mostly related to atmospheric variability. It is apparent that AIFS represents a step change in forecast skill compared to IFS.

\begin{figure}[htpb]
\centering
\includegraphics[trim={100 80 320 100}, clip, width=0.9\linewidth]{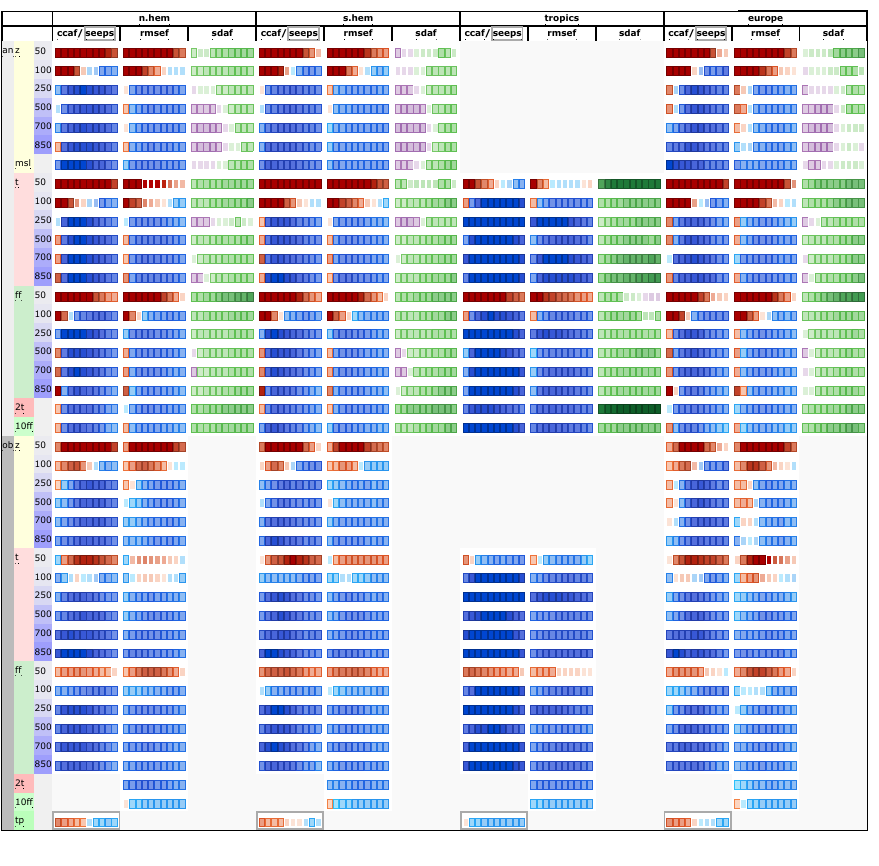}
\caption{Scorecard comparing forecast scores of AIFS versus IFS (2022). Forecasts are initialised on 00 and 12 UTC. Shown are relative score changes as function of lead time (day 1 to 10) for northern extra-tropics (n.hem), southern extra-tropics (s.hem), tropics and Europe. Blue colours mark score improvements and red colours score degradations. Purple colours indicate an increased in standard deviation of forecast anomaly, while green colours indicate a reduction. Framed rectangles indicate 95$\%$ significance level. Variables are geopotential (z), temperature (t), wind speed (ff), mean sea level pressure (msl), 2~m temperature (2t), 10~m wind speed (10ff) and 24 hr total precipitation (tp). Numbers behind variable abbreviations indicate variables on pressure levels (e.g., 500~hPa), and suffix indicates verification against IFS NWP analyses (an) or radiosonde and SYNOP observations (ob). Scores shown are anomaly correlation (ccaf), SEEPS (seeps, for precipitation),  RMSE (rmsef) and standard deviation of forecast anomaly (sdaf, see text for more explanation).}
\label{fig:scorecard}
\end{figure}

Figure~\ref{fig:scorecard} shows a scorecard that compares different forecast skill measures of AIFS to those of IFS across a range of variables. For verification, each system is compared against the operational ECMWF analysis from which the forecasts are initialised. In addition, the forecasts are compared against radiosonde observations of geopotential, temperature and windspeed, and SYNOP observations of 2~m temperature, 10~m wind and 24~h total precipitation. The definition of the metrics, such as ACC (\texttt{ccaf}), RMSE (\texttt{rmsef}) and forecast activity (standard deviation of forecast anomaly, \texttt{sdaf}) can be found in e.g \cite{benbouallegue2023rise}.

AIFS shows better scores throughout the troposphere, up to 100~hPa. Forecast improvements are substantial, of the order of $~10\%$. IFS shows better scores at 50~hPa, which is in line with the reduced weight given to these pressure levels in training. In general, AIFS exhibits substantially lower forecast errors than IFS after day 1 of the forecast. Forecast scores are worse for AIFS compared to IFS at day 1 when forecast data is verified against NWP analyses, e.g. for 850~hPa wind speed in the northern and southern extra-tropics. However, this degradation is not present in verification against radiosonde observations. IFS analyses and forecasts can exhibit correlations that can extend some time into the forecast, and this might be an explanation for the apparent larger errors of AIFS forecasts at day 1. 

Similarly, while AIFS's 100~hPa scores also seem worse than IFS for geopotential ($z$) and temperature, when verified against analyses, this is not necessarily the case when compared against radiosondes. For example, for Northern Hemisphere extra-tropics temperature at 100~hPa, RMSE and ACC are degraded for AIFS compared to IFS in the analysis based verification, while they appear improved when AIFS and IFS are verified against radiosonde observations.

\begin{figure}[htpb]
\centering
\begin{subfigure}{0.48\textwidth}
    \includegraphics[width=\linewidth]{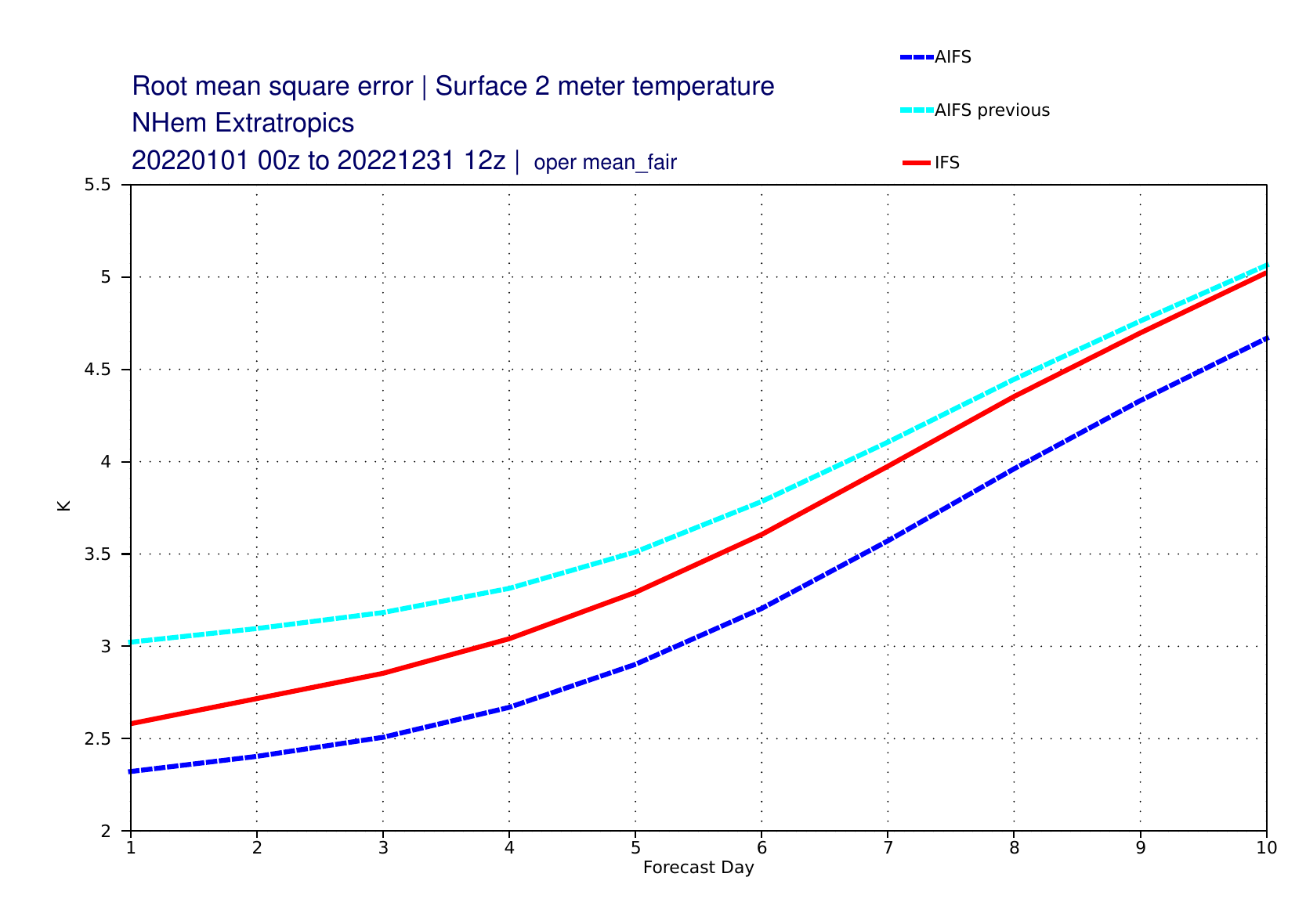}
\end{subfigure}
\hfill
\begin{subfigure}{0.48\textwidth}
    \includegraphics[width=\linewidth]{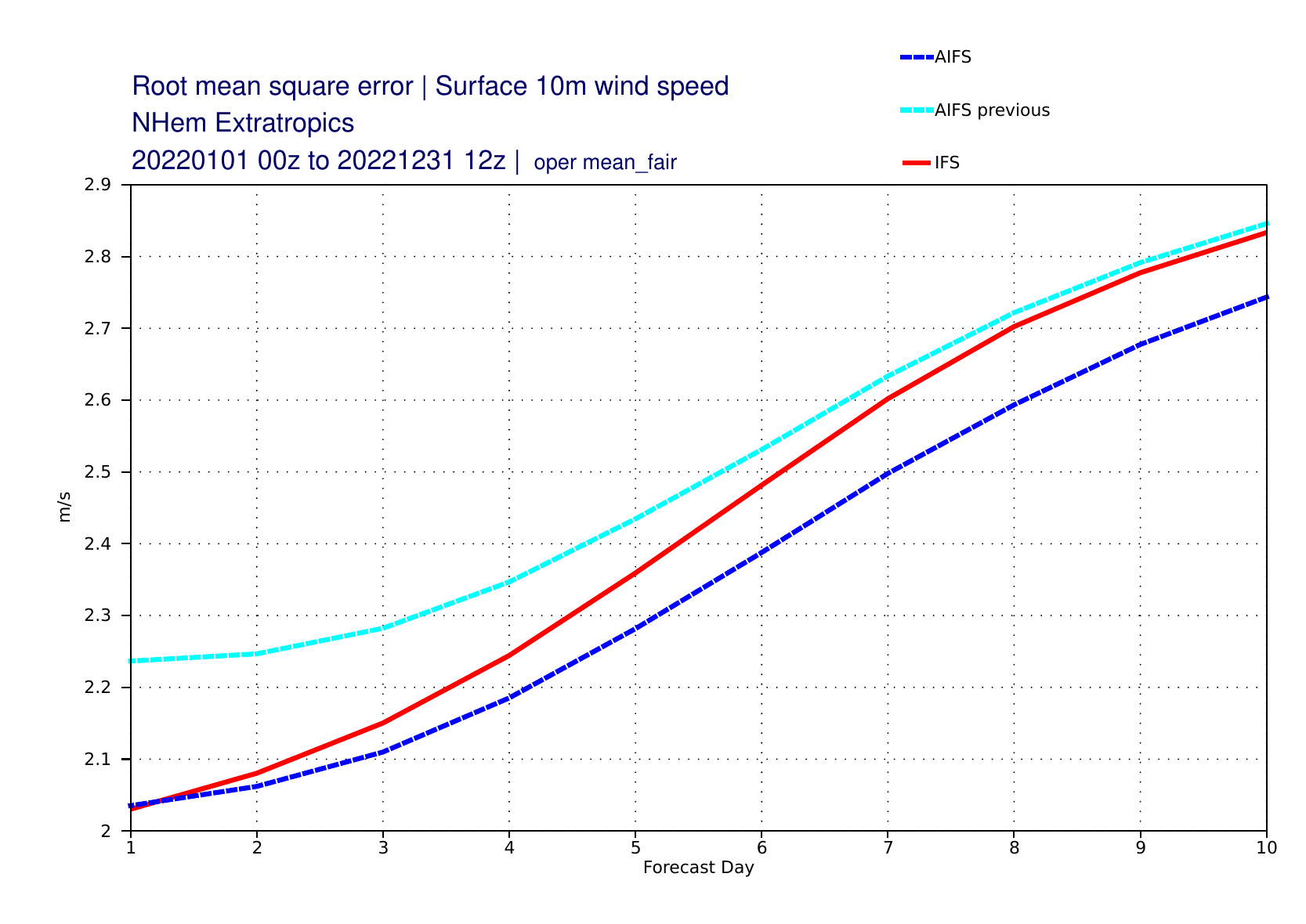}
\end{subfigure}
\caption{RMSE of 2m temperature (left) and 10m wind speed (right) for AIFS (blue, dashed), IFS (red, solid) and previous lower resolution AIFS version (cyan, solid). Forecasts for 2022, Northern Hemisphere. Forecasts are initialised on 00 and 12 UTC each day.}
\label{fig:2t10ff}
\end{figure}

In line with upper-air scores, AIFS surface forecasts score consistently better than IFS when verified against analyses and observations, e.g., 2m temperature and 10m wind as shown in Figure~\ref{fig:2t10ff}. For reference, we also show the initial AIFS version, with approximately 1\degree spatial resolution. While the initial lower resolution AIFS version was actually quite close to the current AIFS in terms of upper-air scores (not shown), large differences become apparent when comparing surface scores against observations. For surface fields, spatial resolution is of great importance.

Total precipitation results are more mixed. In the extra-tropics, AIFS is less skilful than IFS in terms of SEEPS (\cite{rodwellseeps}) at shorter forecast ranges. At longer forecast ranges, AIFS is more skilfull than IFS. In the tropics, total precipitation forecast skill is in general improved compared to IFS.

\begin{figure}[htpb]
\centering
\begin{subfigure}{0.45\linewidth}
    \includegraphics[width=\linewidth]{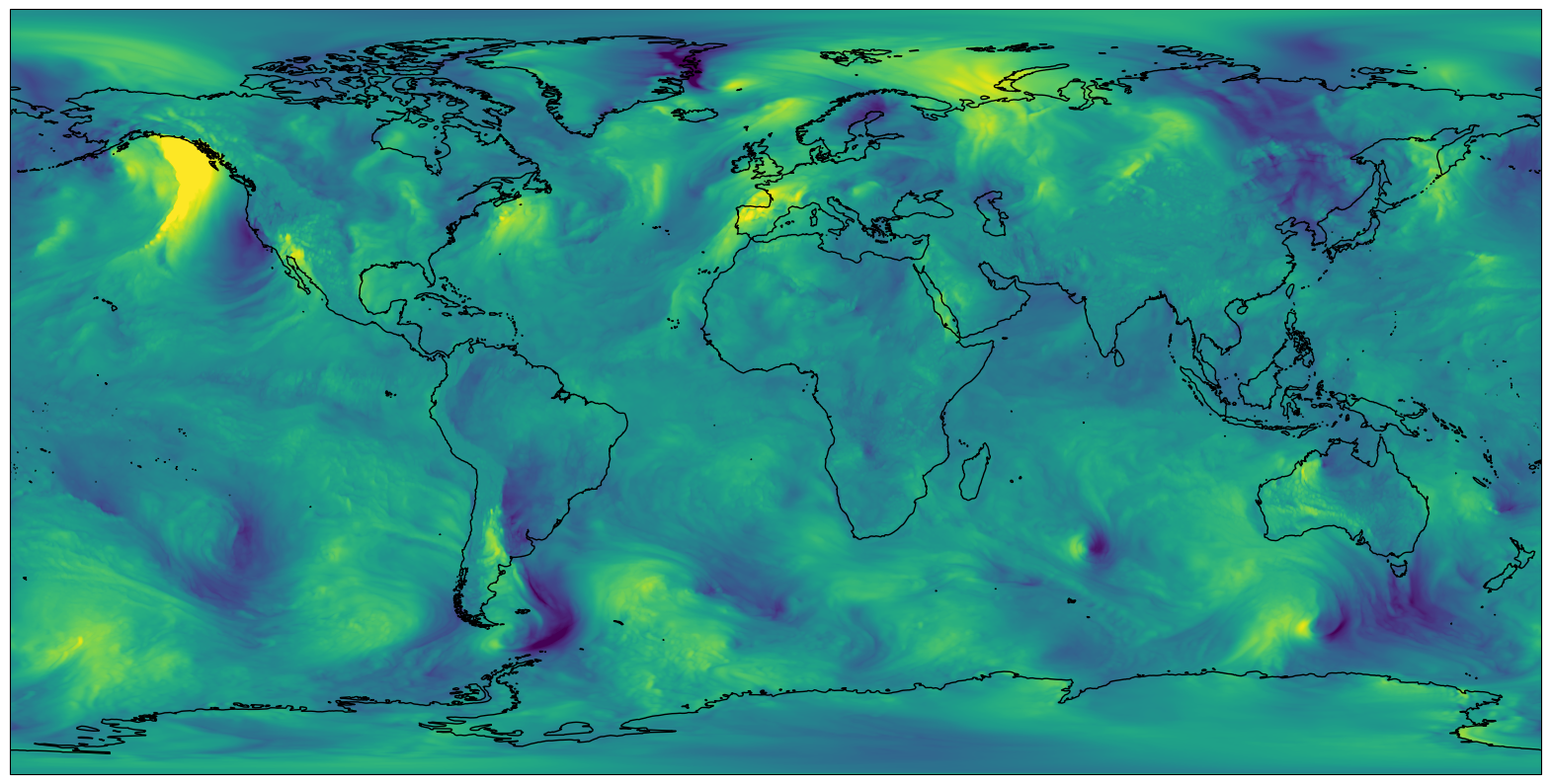}
    \label{fig:12h_ifs}
\end{subfigure}
\hfill
\begin{subfigure}{0.45\linewidth}
    \includegraphics[width=\linewidth]{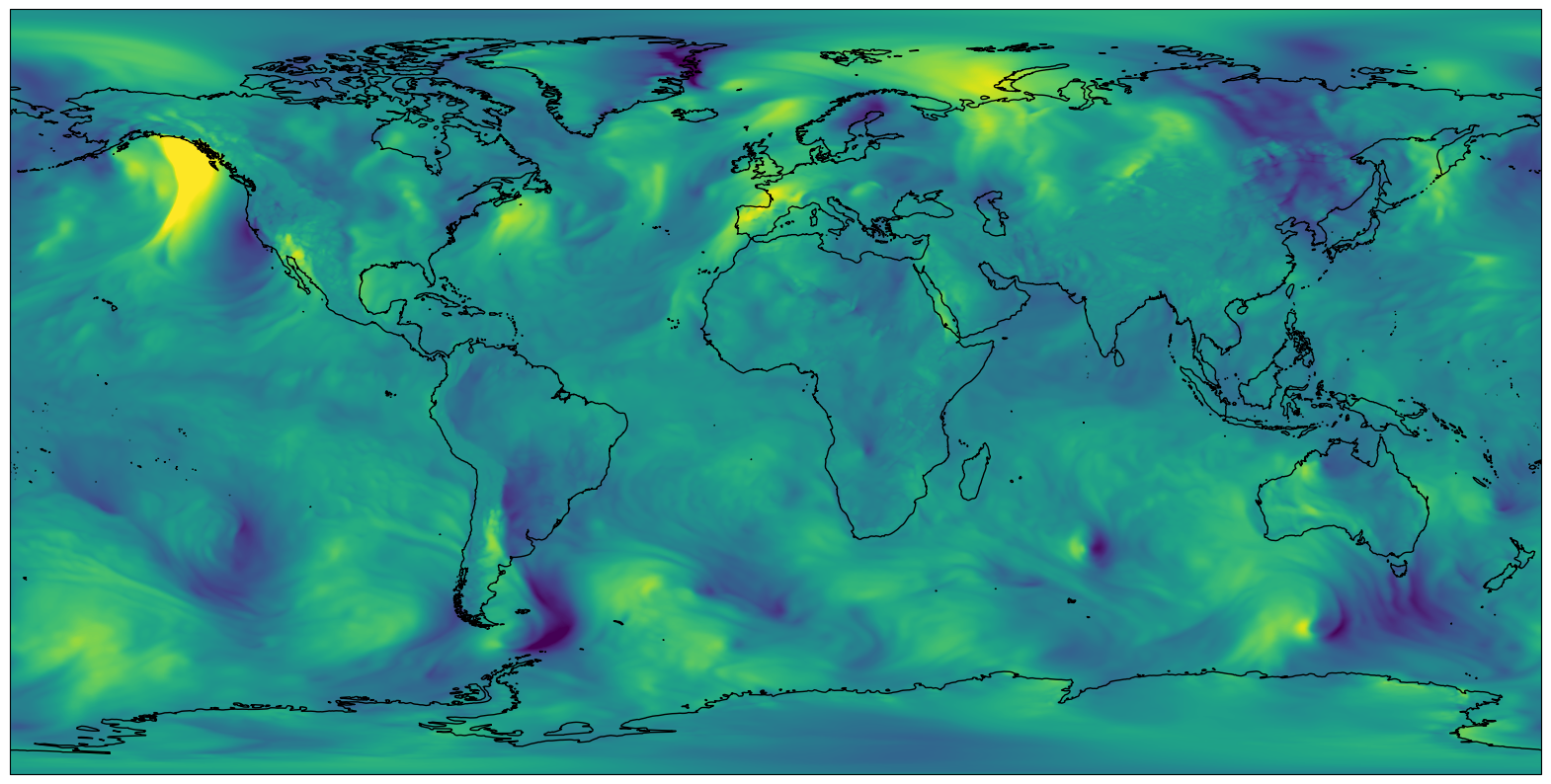}
    \label{fig:12h_aifs}
\end{subfigure}

\begin{subfigure}{0.45\linewidth}
    \includegraphics[width=\linewidth]{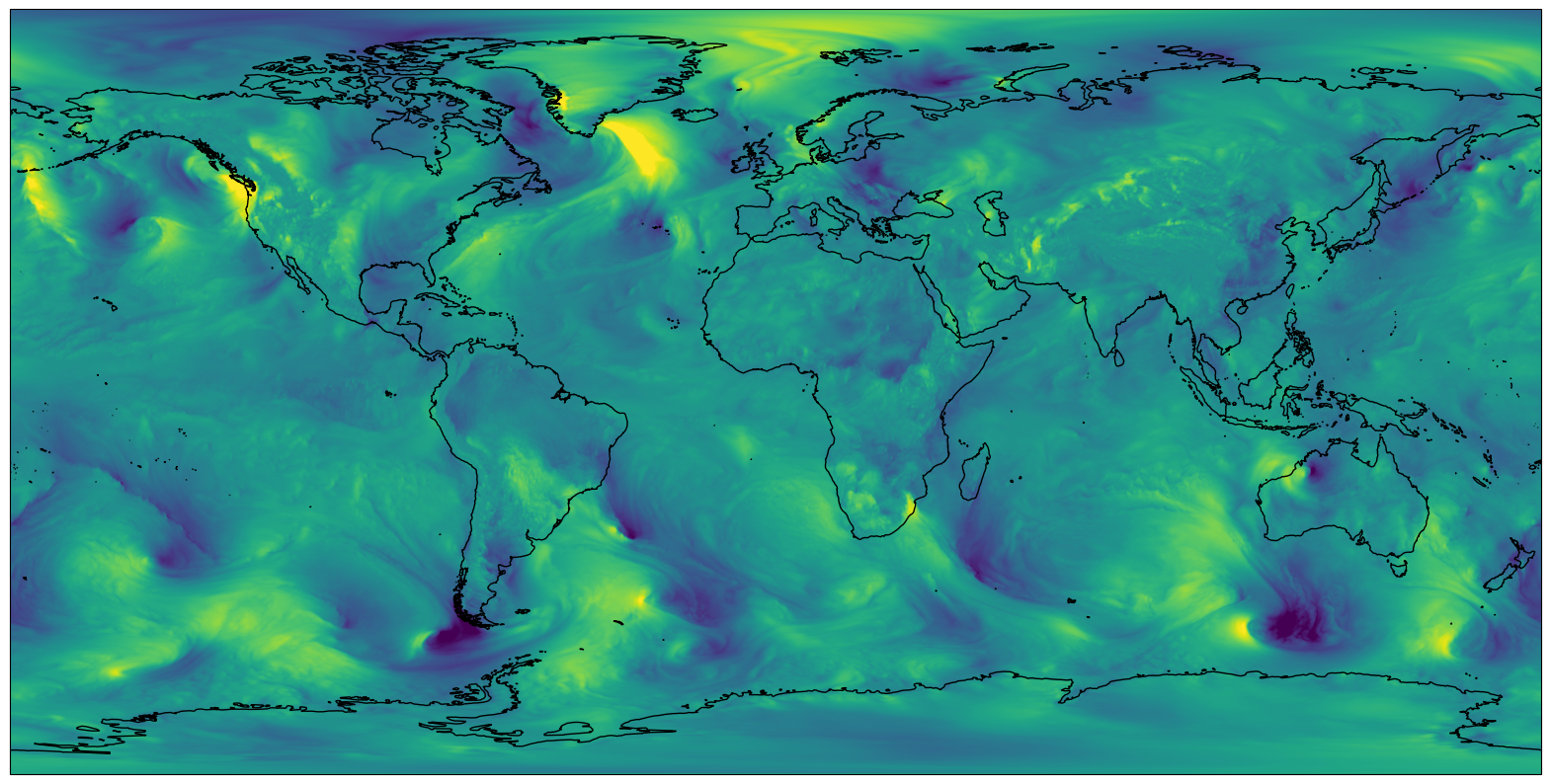}
    \label{fig:120h_ifs}
\end{subfigure}
\hfill
\begin{subfigure}{0.45\linewidth}
    \includegraphics[width=\linewidth]{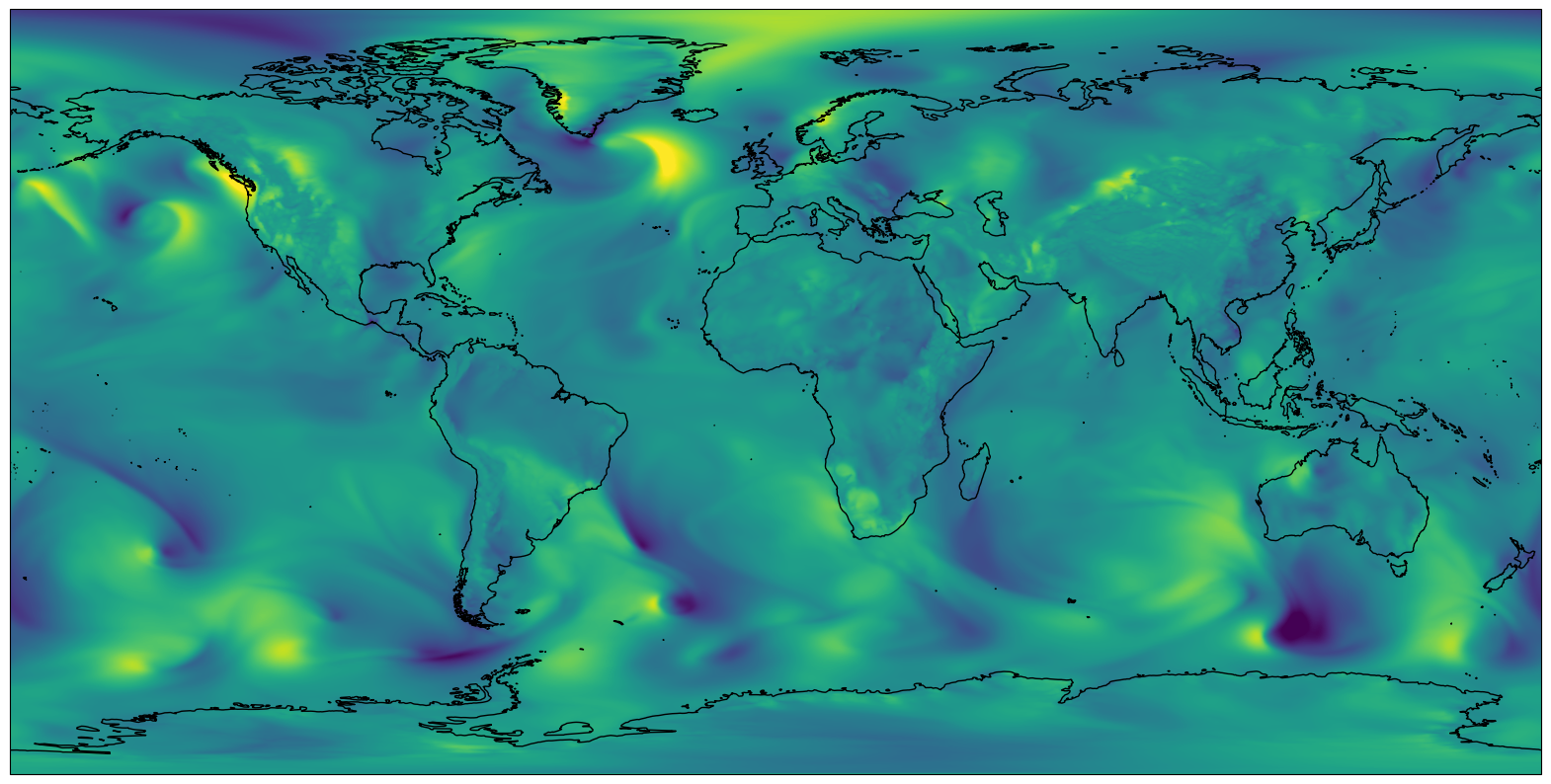}
    \label{fig:120h_aifs}
\end{subfigure}

\begin{subfigure}{0.45\linewidth}
    \includegraphics[width=\linewidth]{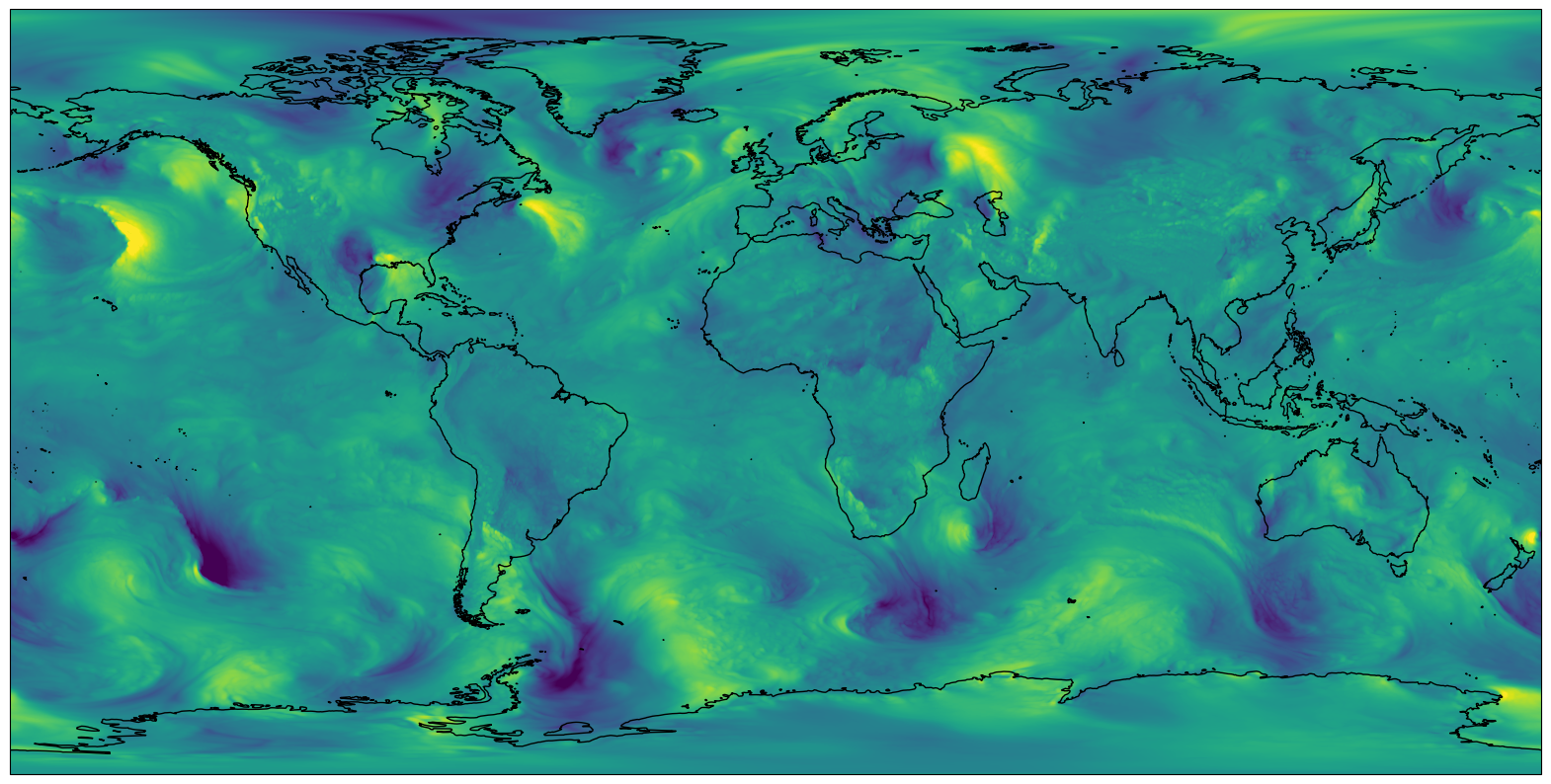}
    \label{fig:240h_ifs}
\end{subfigure}
\hfill
\begin{subfigure}{0.45\linewidth}
    \includegraphics[width=\linewidth]{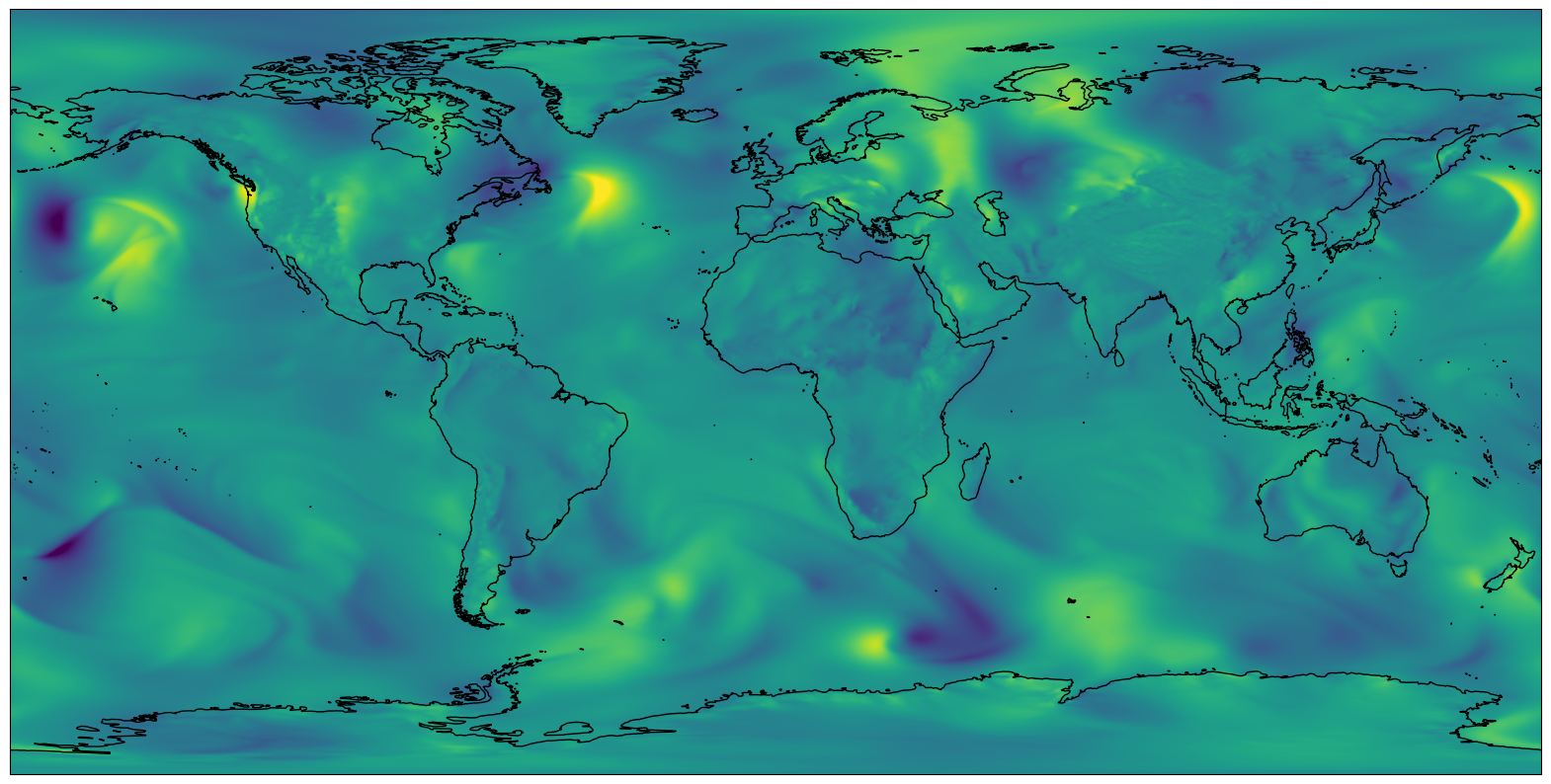}
    \label{fig:240h_aifs}
\end{subfigure}
\caption{Meridional wind at 850~hPa: IFS (left) and AIFS (right) for 1 January 2023 00 UTC date, 12~h (top), 120~h (middle) and 240~h (bottom) forecasts. For plotting, IFS and AIFS forecast fields were interpolated to a 0.25$\degree$ regular latitude-longitude grid. The MSE objective used in AIFS training leads to more smoothing at longer lead times.}
\label{fig:v850}
\end{figure}

The scorecard also shows that overall AIFS's forecast activity is somewhat reduced compared to IFS. The reduction of forecast activity is accompanied by a smoothing, or "blurring", of the forecast fields with lead-time, a behaviour described by, e.g., \cite{pathak2022fourcastnet}, \cite{keisler2022forecasting} and \cite{lam2022graphcast}, which is also visible in global spectra (not shown). While short-range forecasts still contain quite a lot of small scale structures, these are washed out in longer range forecasts (Fig.~\ref{fig:v850}).

\begin{figure}[htbp]
\centering
\includegraphics[width=0.8\linewidth]{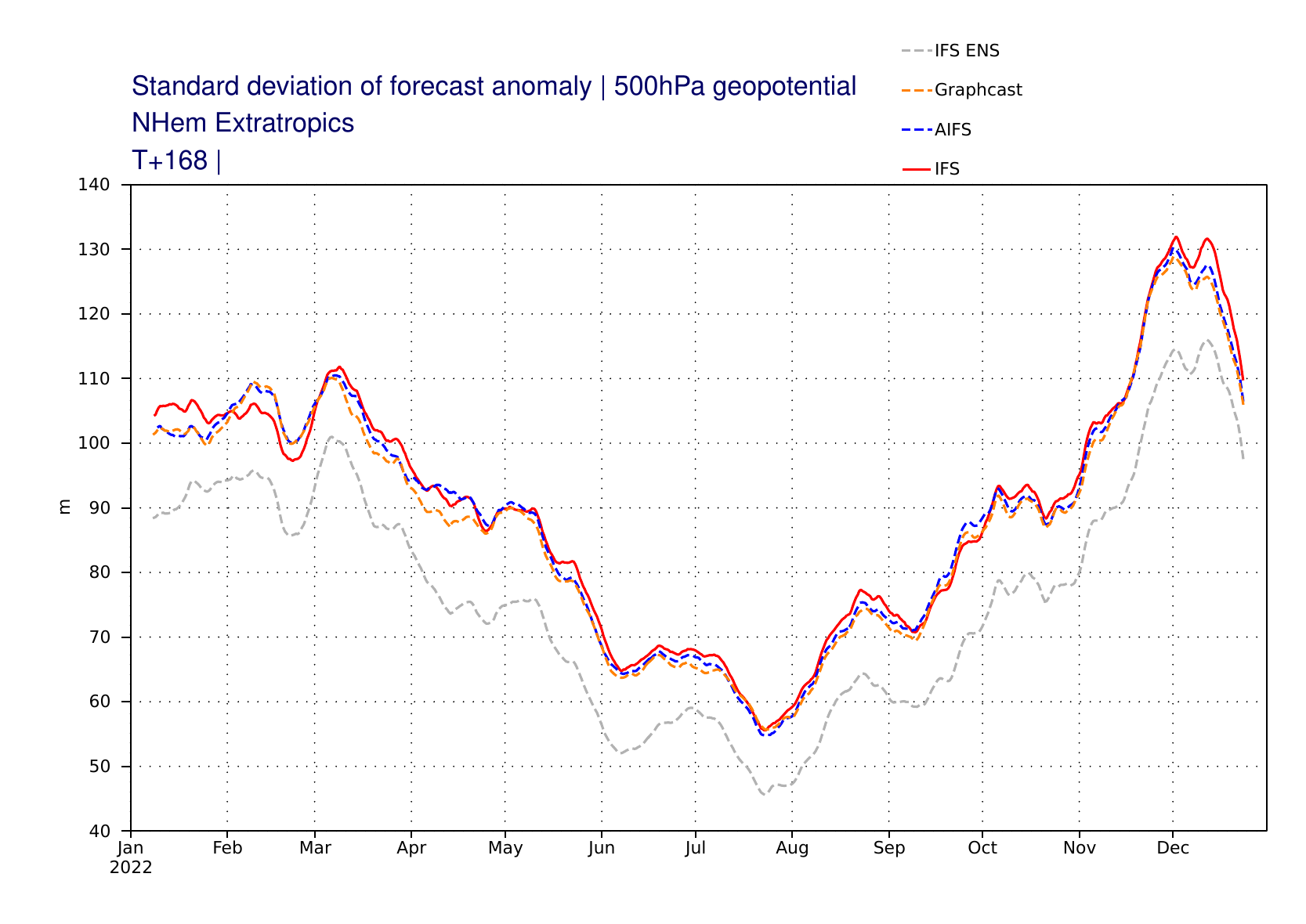}
\includegraphics[width=0.8\linewidth]{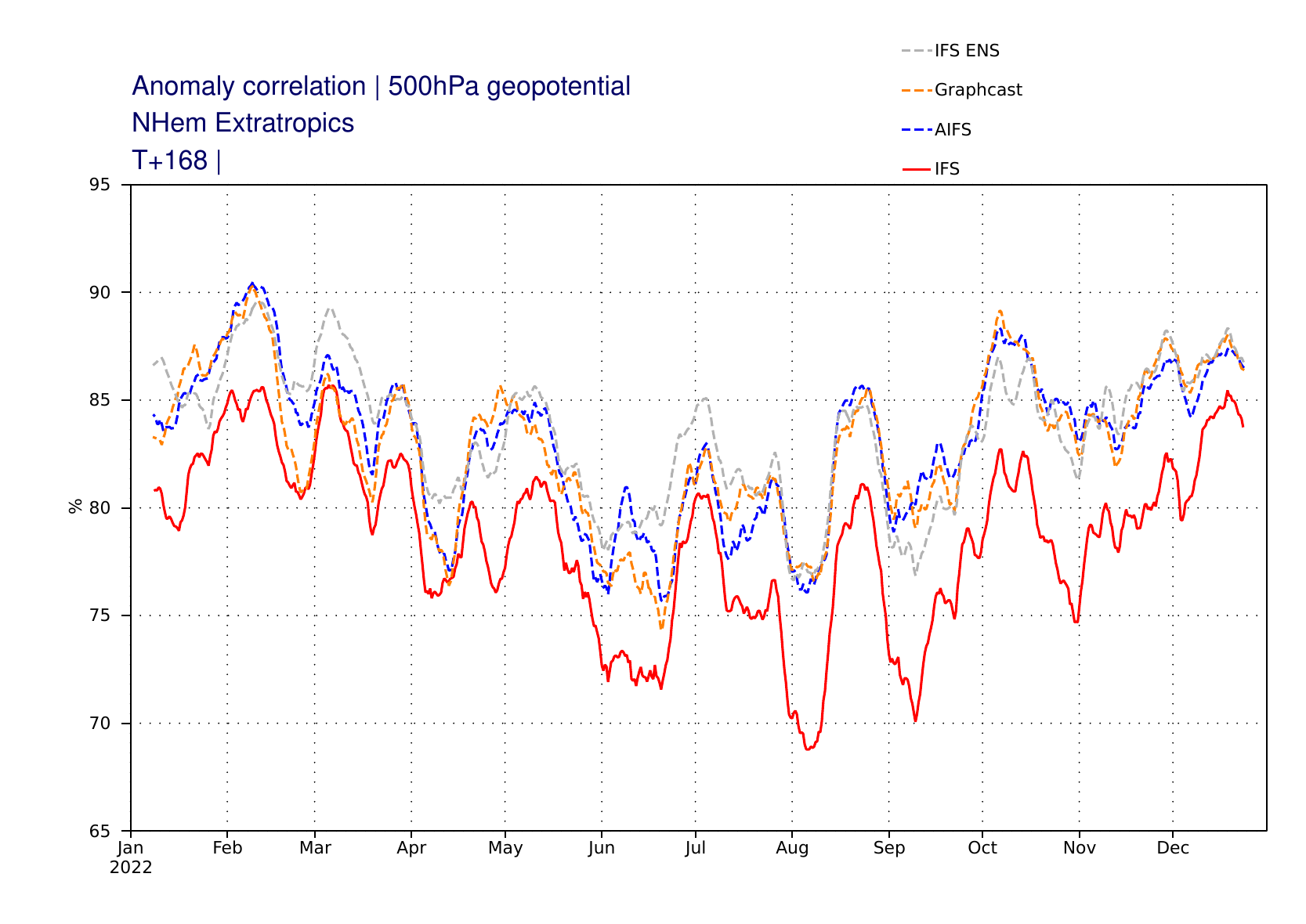}
\caption{AIFS, IFS, Graphcast and IFS ENS mean comparison of 30 day running means of 7 day forecasts. Shown is geopotential at 500~hPa, Northern extra-tropics, 2022, top: forecast activity (standard deviation of forecast anomaly), bottom: ACC (anomaly correlation); AIFS (blue), IFS (red), GraphCast (orange) and IFS-ENS (grey).}
\label{fig:comparison}
\end{figure}

Nevertheless, the forecast activity is not as strongly reduced as in the case of the ensemble mean (compare the blue, red and grey curves in Fig.~\ref{fig:comparison}) of IFS-ENS (ECMWF's NWP ensemble forecasts, \cite{moltenieps}; see \cite{lang2021more}, \cite{lang81380} for latest configuration), which filters out unpredictable features with lead time. 

For reference, we also include the activity of GraphCast forecasts (\cite{lam2022graphcast}), which is quite similar to AIFS. The GraphCast version shown here has been fine-tuned by \cite{lam2022graphcast} to be initialised from operational IFS analyses and is run daily at ECMWF. When looking at forecast performance, it is also apparent that the difference between AIFS and GraphCast is much smaller than between AIFS and IFS. This is in line with both being trained on the same data (ERA5) and IFS analyses, and similarities of their respective training regimes (rollout training, etc.). Interestingly, there are longer periods during the year, where one model seems to perform better than the other, i.e. one curve is below the other for several weeks. This may either be caused by intrinsic variability or by differences in model architectures or training and fine-tuning strategies. It also highlights the need for long verification statistics. 

\begin{figure}[htbp]
    \centering
    \begin{subfigure}[b]{0.42\textwidth}
        \centering
        \includegraphics[width=\textwidth]{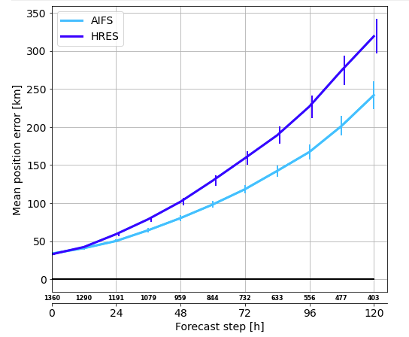}
    \end{subfigure}
    \hfill
    \begin{subfigure}[b]{0.42\textwidth}
        \centering
        \includegraphics[width=\textwidth]{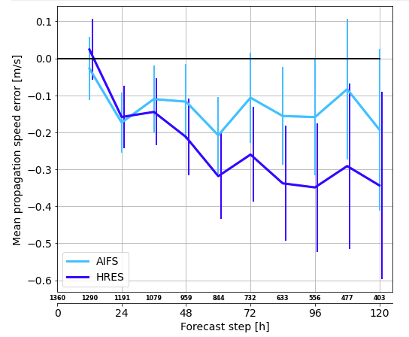}
    \end{subfigure}
    \newline
    \begin{subfigure}[b]{0.42\textwidth}
        \centering
        \includegraphics[width=\textwidth]{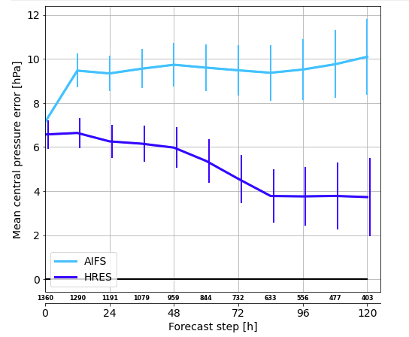}
    \end{subfigure}
    \hfill
    \begin{subfigure}[b]{0.42\textwidth}
        \centering
        \includegraphics[width=\textwidth]{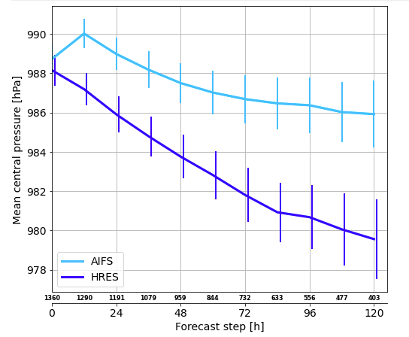}
    \end{subfigure}
    \caption{Comparison of IFS and AIFS Mean tropical cyclone position error (upper-left), mean tropical cyclone propagation speed (upper-right), tropical cyclone mean central pressure error (bottom-left) and tropical cyclone mean central pressure (bottom right) of AIFS (light blue) and IFS (dark blue), January 2022 to December 2023.}\label{fig:TCs}
\end{figure}

In line with results reported in the literature for data-driven models (see e.g., \cite{bi2023accurate}), \cite{lam2022graphcast}), AIFS exhibits substantially lower tropical cyclone (TC) position forecast errors than IFS (see Fig.~\ref{fig:TCs}). To a large extent, this can be explained by a reduced slow bias in the TC propagation speed, which leads to lower along-track TC forecast errors. On the other hand, TC intensity errors are larger than the ones from IFS. This is the case for central pressure, as well as maximum wind speed (not shown). Here, AIFS produces less intense TCs than IFS on average, which manifests in a larger TC intensity bias. This might be explained by the on average too low intensity of TCs in ERA5 and IFS analyses, as well as the tendency of AIFS to smooth forecast fields during the forecast. There is also an indication that, on average, AIFS generates fewer TCs than IFS (not shown).

\section{Discussion and conclusions}
AIFS demonstrates highly competitive forecast performance for both, upper-air and surface variables. These results hold when verified against NWP analyses as well as radiosonde and SYNOP observations. In addition, AIFS produces e.g. accurate tropical cyclone track forecasts.  

The current implementation of AIFS shares certain limitations with some of the other data-driven weather forecast models that are trained with a weighted MSE loss, such as blurring of the forecast fields at longer lead times. The blurring can be traced back to the MSE training objective, where the model learns an implicit smoothing in order to avoid the double penalty (e.g. \cite{Hoffman1995, ebertdouble}) of misplaced structures in the forecast. The smoothing depends on the length of the optimisation window used during training. In the extreme, forecasts produced by a model optimised out to day 10 would be similar to an ensemble mean, with strongly reduced activity and detail (see \cite{blogaifs2}). Preliminary results show that training towards a probabilistic objective is producing sharp forecast fields throughout the forecast, consistent with the results reported by \cite{price2023gencast}.

Further limitations include the relatively ad-hoc approach to fine-tuning on the operational IFS analysis. A more in-depth exploration of different fine-tuning strategies could be beneficial. 

AIFS currently uses "full" normalised states in its training loss. This stands in contrast to predicting normalised forecast \textit{tendencies} as done by, e.g., \cite{keisler2022forecasting} or \cite{lam2022graphcast}. The potential downside here is that the output of the last linear layer of AIFS that maps latent space variables to physical space may not be adequately normalised. This will likely be revised in a future AIFS upgrade. This revision will potentially also facilitate improved strategies for the loss scaling that is currently done empirically on a per variable basis. 

Currently, AIFS exhibits reduced forecast skill in the stratosphere forecast owing to the linear loss scaling with height (e.g. \cite{lam2022graphcast}), i.e., only little emphasis is put on variables in the stratosphere. Optimising loss scaling with height is expected to improve stratospheric forecast skill.
 
A further area for potential improvements is the currently reduced intensity of some high-impact systems such as tropical cyclones. Here, improved input data, such as high resolution re-analyses could potentially help. As stated above, the smoothing with forecast lead time could be alleviated by refining the training loss or by training probabilistic models.

The strength and weaknesses of data-driven forecast models have been investigated in \cite{benbouallegue2023rise} and \cite{Charlton-Perez2024}, and in an idealised context in \cite{hakim2023dynamical}. However, further work is required to assess how the models perform in practice and how they compare to physics-based NWP models.

The modular architecture of AIFS makes it highly flexible, scalable and extendable. Ongoing research work focuses on the extension of AIFS to probabilistic ensemble forecasting, either via CRPS-based training (e.g. \cite{pacchiardi2024probabilistic}, \cite{kochkov2024neural}) or diffusion-based generative models (\cite{DBLP:journals/corr/Sohl-DicksteinW15, ho2020denoising, karras2022elucidating, price2023gencast}).
Because of the comparatively small cost associated with a single data-driven forecast, it will be possible to run ensemble forecasts consisting of a large number of members. 

Other important areas of research are data-driven long-range forecasts, e.g. monthly and seasonal forecasts, improving the representation of precipitation fields (\cite{rainingdatablog2024}) and data-driven regional modelling (\cite{regionalmodelingblog2024}). A further area of exploration is the augmentation of AIFS training and initialisation with observations, e.g. SYNOP and satellite observations, to correct for analysis errors and allow the model to produced improved point forecasts. These developments will also make it possible to generate fully observation-driven (\cite{mcnally2024}) forecasts.

We plan to release the AIFS source code and weights under an open source license later this year, together with the Anemoi framework. AIFS forecasts are run four times daily at ECMWF, and are available to the public under ECMWF's open data policy.

\paragraph{Acknowledgments:} \textit{We acknowledge PRACE for awarding us access to Leonardo, CINECA, Italy.}

\bibliographystyle{plainnat}
\bibliography{refs}

\end{document}